\documentclass{emulateapj}
\usepackage{apjfonts}

\catcode`\@=11
\newcommand{\gapprox}{\mathrel{\mathpalette\@versim>}}
\newcommand{\lapprox}{\mathrel{\mathpalette\@versim<}}
\newcommand{\propapprox}{\mathrel{\mathpalette\@versim\propto}}
\newcommand{\@versim}[2]
  {\lower3.1truept\vbox{\baselineskip0pt\lineskip0.5truept
\ialign{$\m@th#1\hfil##\hfil$\crcr#2\crcr\sim\crcr}}}
\catcode`\@=12

\shorttitle{CHANDRA OBSERVATION OF KEPLER'S SNR}

\begin{document}

\title{A Deep {\sl Chandra} Observation of Kepler's Supernova Remnant:
A Type Ia Event with Circumstellar Interaction
}

\author{Stephen P. Reynolds,\altaffilmark{1}
Kazimierz J. Borkowski,\altaffilmark{1}
Una Hwang, \altaffilmark{2}
John P.~Hughes,\altaffilmark{3}
Carles Badenes,\altaffilmark{3,4}
J.M.~Laming,\altaffilmark{5}
and J.M.~Blondin\altaffilmark{1}}

\altaffiltext{1}{Department of Physics, North Carolina State University,
  Raleigh NC 27695-8202; stephen\_reynolds@ncsu.edu} 
\altaffiltext{2}{NASA/GSFC, Code 660, Greenbelt, MD 20771}
\altaffiltext{3}{Department of Physics and Astronomy, Rutgers University, 
Piscataway, NJ 08854-8019}
\altaffiltext{4}{{\it Chandra} Fellow.}
\altaffiltext{5}{NRL, Code 7674 L, Washington, DC 20375-5320}

\begin{abstract}

We present initial results of a 750 ks Chandra observation of the
remnant of Kepler's supernova of AD 1604.  The strength and prominence
of iron emission, together with the absence of O-rich ejecta,
demonstrate that Kepler resulted from a thermonuclear supernova, even
though evidence for circumstellar interaction is also strong.  We have
analyzed spectra of over 100 small regions, and find that they fall
into three classes.  (1) The vast majority show Fe L emission between
0.7 and 1 keV and Si and S K$\alpha$ emission; we associate these with
shocked ejecta.  A few of these are found at or beyond the mean blast
wave radius.  (2) A very few regions show solar O/Fe abundance
rations; these we associate with shocked circumstellar medium (CSM).
Otherwise O is scarce.  (3) A few regions are dominated by continuum,
probably synchrotron radiation.  Finally, we find no central point
source, with a limit $\sim 100$ times fainter than the central object
in Cas A.  The evidence that the blast wave is interacting with CSM
may indicate a Ia explosion in a more massive progenitor.

\end{abstract}


\section{Introduction}

Kepler's supernova of 1604 \citep{kepler1606} was the most recent
well-observed Galactic supernova.  The supernova type has been
controversial for decades; the Galactic latitude ($b = 6\fdg8$)
implies a distance above the Galactic plane of 470 pc for a distance
of 4 kpc \citep{sankrit05}, arguing for a Type Ia (Population II)
progenitor, but optical observations \citep{blair91} show radiative
shocks in dense knots, suggestive of circumstellar material (CSM) shed
by the progenitor.  Bandiera (1987) proposed that the progenitor was a
massive runaway star with a strong wind; Kepler's bright N rim
suggests a bow shock directed away from the Galactic plane.

X-ray observations of young supernova remnants can give powerful 
constraints on progenitor models and explosion mechanisms; whether
Kepler had a Type Ia or core-collapse origin, the detailed distribution
of ejected material and the composition of CSM could be used to test
SN models.  We undertook a long observation of Kepler with the {\sl
Chandra} X-ray Observatory to study the composition and distribution
of ejecta and CSM, to search for any compact object, and to search
for convincing evidence of the SN type.

\section{Observations and Results}

We observed Kepler's SNR for 741 ks with the {\sl Chandra} X-ray
Observatory ACIS-S CCD camera as a Large Project in six installments
between April and July 2006.  In all six, Kepler was positioned
(slightly differently) on the S3 back-illuminated chip.  Data were
processed using CIAO v3.4 and calibrated using CALDB v3.1.0.  A large
background region to the N of the remnant (covering most of the
remaining area on the S3 chip) was used for almost all spectra except
when local backgrounds were called for, as described below.  Spectral
analysis was performed with XSPEC v.12 \citep{arnaud96}. We used
the nonequilibrium-ionization (NEI) v2.0 thermal models, based on the
APEC/APED spectral codes \citep{smith01} and augmented by addition of
inner-shell processes \citep{badenes06}.

We obtained about $3 \times 10^7$ source counts, with fewer than 3\%
background (though some of those may be dust-scattered source
photons).  S/N permitting, spectra were extracted between 0.3 and 8
keV.  Images were smoothed using platelets with a default smoothing
parameter $\gamma = 1/2$ \citep{willett06}.

\begin{figure}
\centering
\epsscale{1.1}
\includegraphics[width=3truein]{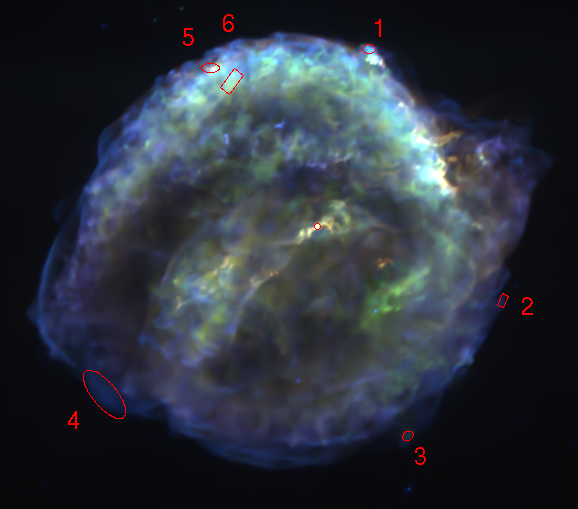}
\caption{Merged image between 0.3 and 8 keV.  Red:  0.3 -- 0.72 keV;
green, 0.72 -- 1.7 keV; blue, 1.7 -- 8 keV.  All three images were
smoothed using platelet smoothing \citep{willett06}.  Image size
is $4\farcm7 \times 3\farcm9$.  Regions from which spectra of Figs.~\ref{spectra1}
and \ref{spectra2}
were extracted are shown.
\vspace{-0.1truein}
\label{totx}
}
\end{figure}

Figure~\ref{totx} shows the full observation, with red for energies
between 0.3 and 0.72 keV (including primarily O), green for 0.72 --
1.7 keV (primarily Fe L-shell transitions, also including Ne and Mg
K$\alpha$ complexes), and blue for 1.7 -- 8 keV (with a strong
contribution from Si K$\alpha$).  Immediately evident is the absence
of dominantly soft-spectrum regions, and the hardness of emission at
the periphery.  A few regions do show reddish colors; representative
spectra are discussed below.

No central point source is obvious.  The bright central knot is
resolved, and several fainter central knots, though perhaps
unresolved, also have spectra of shocked gas. A faint source could be
concealed under bright extended emission.  Our conservative estimate
assumes a high total background of 860 cts per ACIS S3 pixel,
appropriate for the bright central band of emission. A 5$\sigma$
detection of a point source within the 80\%\ encircled power radius of
$0\farcs685$ (6 px area) would require 400 cts, or $6 \times 10^{-4}$
cts s$^{-1}$. The compact central object in Cas A \citep{pavlov00}, if
placed within Kepler and using the absorbing column density determined
below, would produce 0.18 cts s$^{-1}$. We conclude that we would
detect such a source even if its luminosity were $\sim 100$ times
fainter.

For spectral analysis, we fixed the absorbing column density at $N_H =
5.2 \times 10^{21}$ cm$^{-2}$, a value we found by fitting lineless
regions in the SE with an absorbed synchrotron model {\tt srcut} using
\citet{wilms00} abundances.  We used the three-color image of
Fig.~\ref{totx} to aid in identifying small, relatively homogeneous
regions of contrasting spectral character.  We identified and examined
about 100 regions.  The vast majority show prominent lines of Fe
L-shell transitions, He-like Si, S, Ar, and Ca, and Fe K$\alpha$, and
a handful had contrasting properties.  We found that the spectra can
be separated into three main classes.  Over 90\% of the regions showed
Fe- and Si-rich spectra; we identify these with SN ejecta.  A few
regions showed spectra with O and Mg emission consistent with solar-
or near-solar abundances; we identify these with CSM.  Finally, a few
regions showed continuum-dominated spectra.

\begin{figure}
\centering

\includegraphics[width=3truein]{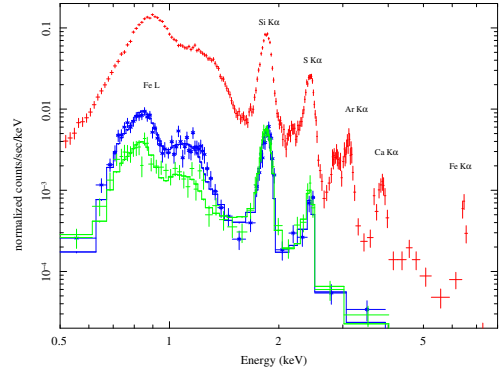}
\caption{Spectra of three outer knots shown on Fig.~\ref{totx}.  Top
(red): Bright Si- and S-rich knot 1 in the N. Fe L shell, Fe K$\alpha$, 
Ar, and Ca emission are also strong.  Middle (blue) and
bottom (green): Knots in the S (3) and W (2), with plane shock model fits
shown. 
\label{spectra1}
}
\end{figure}

\begin{figure}
\centering

\includegraphics[width=3truein]{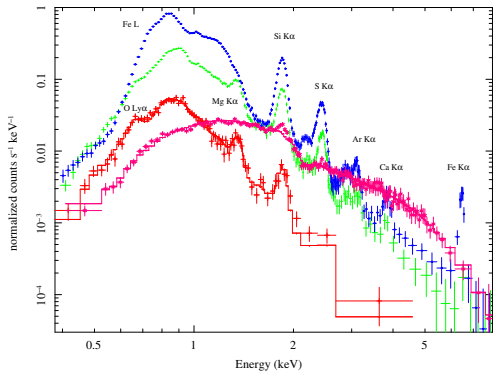}
\caption{Spectra of four knots shown on Fig.~\ref{totx}.
Top (blue):  Fe L region 6.  No O or Mg features are evident.
Second from top (green):  Soft-spectrum knot 5.
An inflection at 0.6 keV is
O Ly$\alpha$, and a feature at Mg K$\alpha$ is labeled.
Third from top at 0.8 keV (red):  bright central knot.  Note O and Mg features.
Bottom at 0.8 keV (magenta): synchrotron-dominated region 4 at rim in SE.
Only a faint Si feature is apparent.
\label{spectra2}
}
\end{figure}

Examples of ejecta spectra (regions 1 -- 3 on Fig.~\ref{totx})
are shown in Fig.~\ref{spectra1} and the
top spectrum (Fe L region, 6) in Fig.~\ref{spectra2}.  In particular,
Fig.~\ref{spectra1} shows spectra of three outer knots, where we might
expect to find the lightest nucleosynthetic products at the largest
distances, specifically the O that should indicate a CC explosion.
All of them, however, show strong Si, S, and Fe L emission; prominent
Ar, Ca, and Fe K$\alpha$ lines are also present in the bright N
knot 1. The western and southern knots 2 and 3 have comparable Si and S
emission, but they differ in their Fe L-shell emission.

Knot 3 is well modeled by a heavy-element plane shock
\citep{hendrick} with velocity $v_s$ of 1010 km s$^{-1}$ ([889, 1120]
km s$^{-1}$ 90\%\ confidence interval), shock ionization age $\tau =
2.69 (2.32,3.18) \times 10^{10}$ cm$^{-3}$ s, Si/Fe and S/Fe ratios
(by number) of 2.64 (2.35,2.92) and 2.24 (1.59,2.87), respectively.
The knot is clearly composed of ejecta.  A trace amount of Mg (Mg/Fe
$=0.052$ [0.003,0,10]) has been included, because it formally improves
the fit, but its presence is much in doubt in view of deficiencies in
Fe L atomic data. A synchrotron component (constrained by the absence
of radio flux from the knot [DeLaney et al.~2002]) improves the fit
quality.
We assumed no collisionless heating at the shock; if such heating were
present, $v_s$ could be substantially less.  Si and S are the primary
constituents of the southern knot, but a significant ($1/4$ by mass)
amount of Fe is present.  

In the western knot 2, the shock speed is lower ($v_s = 500 [360,830]$
km s$^{-1}$) and the ionization age is longer ($\tau = 4.8 [2.7,14]
\times 10^{10}$ cm $^{-3}$ s) but most significant are much higher
Si/Fe and S/Fe ratios, 11.6 (8.9,14.5) and 14.7 (10.2,21.1),
respectively.  There is noticeably less Fe (6\%\ by mass) in this
knot; different Fe contents of knots 2 and 3
partially account for their different Fe L-shell emission strengths.
Finally, in knot 1 the relative strengths of Si, S, and Fe
L-shell emission are similar, so this is also a Si- and S-rich knot
containing Fe. (Because a simple plane shock model failed to provide a
satisfactory fit at high energies, we have not attempted a detailed
abundance analysis for knot 1).

The striking lack of obvious O features in the vast majority of the
regions we examined led us to search specifically for soft-spectrum
regions that might be expected to have a different composition.  We
found a few (all showing a pink color in Fig.~\ref{totx}) with
discernible O features; two are shown in Fig.~\ref{spectra2}.  They
constitute our second class of spectral region.  The central knot has
optical and IR counterparts \citep*{blair91,blair07}, implying that it
is composed of ambient CSM.  We obtained the spectrum shown in
Fig.~\ref{spectra2} after subtracting a local background from the
bright E-W band nearby.  In the plane-shock model fit shown in
Fig.~\ref{spectra2}, we allowed the abundances of Ne, Mg, Si, and Fe
to vary (with S and Ni tied to Si and Fe, respectively). The poor
spectral resolution of CCD's and the uncertain calibration at low
energies, combined with deficiencies in Fe L atomic data, do not allow
us to determine reliable absolute abundances with respect to H and He,
or even a relative N abundance, but supersolar abundances for all
elements except N are ruled out.  We assumed near solar composition by
setting the O abundance to its solar value, and determined relative
abundances with respect to O.  We assumed a supersolar \citep[by a
factor of 3;][]{blair07} N abundance.  We obtained a temperature $kT =
0.64$ keV (0.59,0.70), $\tau = 2.7 (1.8, 3.6) \times 10^{11}$
cm$^{-3}$ s, and abundances with respect to solar of Ne = 0.64 (0.41,
0.91), Mg = 0.83 (0.61, 1.1), Si (tied to S) = 1.4(1.0, 1.9), and Fe
(tied to Ni) = 1.1 (0.94, 1.4). When N was allowed to vary, the
best-fit N abundance is 3.2 (1.9, 4.6), where errors do not include
systematic uncertainties particularly important at low photon
energies. The model underestimates the continuum near the Si and S
K$\alpha$ lines, most likely because of the presence of
multitemperature plasma in the central knot, so abundances of these
elements may be somewhat overestimated. The central knot spectrum is
consistent with the same N-enhanced near-solar abundance composition
found from optical spectra of the CSM in Kepler.
Fig.~\ref{spectra2} also shows another soft-spectrum knot, 5; its
spectrum is consistent with that of the central knot.

Finally, a few regions, primarily in the SE, showed featureless or
almost featureless spectra (one example, region 4, is shown in
Fig.~\ref{spectra2}).  These are well-described by models of
synchrotron emission extending from the radio \citep{reynolds07}; they
form our third class of region, and will be described more fully in a
future publication.

\section{Kepler's progenitor type}

Spectral regions 1, 2, 3, and 6 in our first, Si- and Fe-rich class
are almost certainly ejecta, of a composition completely inconsistent
with a CC event, but expected for a thermonuclear explosion.  In fact,
they resemble Si- and Fe-rich compact ejecta knots in the vicinity of
the eastern blast wave in Tycho \citep{decourchelle01}.  Ejecta knots
far ahead of the blast wave with such an advanced nuclear burning
composition have not been found so far in any CC SNRs, even Cas A;
except for the narrow jet, O is instead the most common element in the
outermost ejecta \citep{fesen06}.  Only about 5\% of our regions fell
into the second class of roughly solar O/Fe and perhaps elevated N, as
expected for CSM.  We can state with confidence that nowhere in Kepler
are there spectral regions consistent with the expected CC O/Fe ratio
of 70 \citep[e.g.,][]{iwamoto99}; it is very difficult to locate O
spectral features at all.  Where strong Fe and Si indicate ejecta, O
is absent.  However, the dense CSM shown in optical observations and
now identified in a few regions in X-rays is not expected for a Type
Ia progenitor.  In view of the unusual nature of our result, we
collect the evidence on Kepler's progenitor type.

{\bf 1. Integrated X-ray spectrum and ejecta.} The integrated spectrum
(which resembles region 6, Fig.~\ref{spectra2} top), as well as most
of our spectral regions, shows very strong Fe L, no apparent O or Mg,
strong Si and often strong Fe K emission.  These elements are all
identified with ejecta.  (O and Mg lines can be seen in the RGS
spectrum; Ballet 2004.)  Even knots 1, 2, and 3 at the farthest
periphery show evidence of strong nuclear processing.  The spectra
resemble the integrated spectrum of the Ia remnant of Tycho's SNR
(Badenes et al. 2006), but are totally unlike that of the young O-rich
CC remnant, 1E0102-70.4, which is dominated by O and Ne emission lines
(Sasaki et al.~2006).  Previous work with {\it ASCA} and {\it
XMM-Newton} (Kinugasa \& Tsunemi 1999, Cassam-Chena\"{i} et al. 2004),
derived substantial masses of shocked Fe, 0.1-0.3 M$_\odot$
\citep{kinugasa99}.

{\bf 2. Scarcity of O.}  By contrast, there is little evidence for O
in Kepler's SNR.  It is seen in only a few regions with spectra quite
different from the ejecta knots in having relatively weak Fe L
emission.  Most of these regions are coincident with optical knots
which have been identified with CSM of roughly solar abundances except
for elevated N.  The X-ray knots do not require enhanced N abundance,
but are consistent with the optical abundances.  In particular, O/Fe
is roughly solar.  In SN ejecta of Type Ia, O/Fe is expected to be
only 0.3-0.7 by number, whereas in CC explosions, it is about 70
(e.g., Iwamoto et al. 1999).  None of the spectra we examined had the
high O/Fe ratios required for CC ejecta.  (Several poorly-understood
luminous Type Ib/c SNe listed by \cite{richardson06} could be an
exception, because their O/Fe ratios are much less than typical for CC
explosions. Without detailed explosion models for these SNe confirming
their CC origin, we cannot exclude such possible extreme CC events on
the basis of the O/Fe ratio alone, but they are ruled out by arguments
below.) We conclude that regions with detectable O are CSM.

{\bf 3. Central source.} No central point source is seen, with a 
limit at least 100 times fainter than the source in Cas A.

We find the X-ray evidence for Kepler's origin in a thermonuclear
event to be compelling.  However, additional evidence can be found
from optical observations.

{\bf 4. Balmer-dominated shocks.} \cite{blair91} discovered
substantial emission from nonradiative, Balmer-dominated shocks.  Such
shocks demand the presence of partially neutral upstream material,
ruling out most CC scenarios.  A hot (i.e., compact) progenitor of a
stripped CC Type Ib/c SN would have a strong ionizing flux before the
explosion, while a cooler star would be larger at the time of the
explosion, resulting in a large ionizing flux at shock breakout. The
presence of neutral H sets an upper limit to the progenitor radius
$R_*$ of $26 \alpha^{0.575} \left(E/10^{51} \ {\rm erg}\right)^{-0.23}
\left(M_{CSM}/M_\odot\right)^{0.575}
\left(M_{ej}/M_\odot\right)^{0.25} R_\odot$ \citep{chevalier05}, where
$13.6\alpha$ eV ($\alpha>1$) is the mean energy of an ionizing photon,
and $M_{CSM}$ and $M_{ej}$ are the shocked CSM and ejecta
masses. Since $M_{CSM}$ is only $\sim 1 M_\odot$ \citep{blair07},
explosions of red supergiants with $R_* \sim 600 R_\odot$ are ruled
out, including not only Type IIP and Type IIL SNe with full and
partially-stripped H envelopes, respectively, but also Type IIb SNe
like 1993J with only a residual ($0.2 M_\odot$) H envelope left on its
He core \citep{woosley94}.  While an exotic stripped CC event,
resulting from a star with a residual hydrogen envelope of low ($\ll
0.2 M_\odot$) mass but large radius, might avoid ionizing the CSM
before or during the explosion, such events become even less probable
than a Type Ia with CSM interaction. An additional problem with a
stripped CC progenitor is a large expected mechanical luminosity of
its stellar wind, since radio observations of circumstellar
interactions in Type Ib/c SNe are consistent with WR winds
\citep[e.g.,][]{chevalier07}. A powerful WR-like wind would have
accelerated and swept away any dense CSM to distances of many pc from
the SN, in obvious conflict with observations.

{\bf 5.  Light curve.} \cite{stephenson02} summarize the historical
data.  Kepler's observed peak magnitude of $m \sim -3$ gives $M_V
\cong -18.8$ (using the observed $E_{B - V} = 0.9$ to obtain $A_V =
2.8$), brighter than most CC events but typical for SNe Ia
\citep[e.g.,][]{hamuy96}.  The light curve, though
resembling a normal Ia, could also describe a SN IIL (but see above).
SN 1987A-like events (fairly compact blue stars with massive H envelopes)
produce light curves that are far too broad, and are subluminous as
well.

\section{Discussion and conclusions}

We believe that X-ray and optical evidence points compellingly to the
conclusion that Kepler resulted from a thermonuclear event.  However,
our deep {\it Chandra} observations also confirm that Kepler's blast
wave is encountering material with at least solar metallicity,
coincident with optically emitting N-enhanced material -- most likely
circumstellar material lost by a fairly massive star, either the
progenitor of the white dwarf or the companion, or both.  Interaction
with CSM is unexpected for Type Ia events.  Some Type Ia SNe have
shown signs of CSM interaction in the form of narrow H lines, like SN
2002ic \citep{hamuy03} and SN 2005gj \citep{aldering06}, but the
interpretation of these objects remains controversial (Benetti et
al.~2006; but see Prieto et al.~2007).  Patat et al.~(2007) give
spectroscopic evidence that SN 2006X is a normal Type Ia SN with CSM
interaction.  Evidence for CSM interaction in SNRs appears to be
relatively more widespread, and in some cases can be associated with
Type Ia events.  Prominent Fe emission in DEM L238 and DEM L249, two
LMC SNRs, indicates a Type Ia origin.  However, the brightness and
location of that Fe, at small radii with a large ionization age,
cannot be explained with standard Type Ia explosions into a uniform
ISM (Borkowski et al 2006); interaction with a nonuniform CSM seems to
be required. Relatively bright central emission dominated by Fe is
common in LMC SNRs \citep{nishiuchi01,vanderheyden04}, suggesting that
this class of SNRs is not a negligible population.  N103B
\citep{lewis03} also shows strong Fe and Si emission and evidence for
CSM interaction.  Kepler's SNR could just be the nearest example of a
class of thermonuclear supernovae interacting with dense CSM from the
progenitor system.

The required CSM might indicate an origin in a younger, more massive
population.  Three recent studies of extragalactic SNe Ia
\citep{mannucci06,sullivan06,aubourg07} propose the presence of a
``prompt'' population (delay $\lapprox 5 \times 10^8$ yr) in addition
to an ``old'' population (delay $\sim$ Gyr) in later-type galaxies.
If stars with initial masses less than about 8 $M_\odot$ produce white
dwarfs, the minimum latency for a thermonuclear supernova would be
only about 50 Myr.  A substantial number of remnants of Type Ia
supernovae should come from the ``prompt'' population.

However, Kepler's origin remains problematic.  The elevated CSM
abundances rule out a very old population, but the systemic velocity
\citep[$> 200$ km s$^{-1}$;][]{blair91} seems to indicate either a
halo population, inconsistent with the abundances, or a runaway,
probably inconsistent with a binary system.  (The Galactic bulge
population would not extend to Kepler's Galactocentric distance of 4.5
kpc.)  The total CSM mass of $\gapprox 1 M_\odot$ \citep{blair07}
requires too large an initial mass for an extreme halo population.
Could Kepler have resulted from the thermonuclear explosion of a
single star?  Current conventional wisdom requires that all AGB stars
eject most of their envelopes to leave a sub-Chandrasekhar white
dwarf; if that ejection were occasionally impeded, a single-star
Type Ia supernova might result \citep[e.g.,][]{tout05}. If Kepler's
progenitor were a single star liberated from the Galactic plane by the
core-collapse explosion of its companion about 3 Myr ago, its maximum
lifetime would only be of order 50 Myr, also arguing for a more
massive progenitor.

Our conclusion that Kepler resulted from a Type Ia event with CSM
interaction has important implications.  Models of SN Ia progenitor
systems make widely varying predictions for the nature of any CSM
\citep[e.g.,][]{badenes07}; these predictions may now be testable.
Similarly, SN Ia explosion models can now be tested in detail against
the abundances and spatial distribution of elements in Kepler, and can
be contrasted with Tycho's SNR, apparently the result of a
conventional SN Ia encountering only low-density, more or less uniform
ISM \citep{badenes06}. The existence of a more massive subclass of SNe
Ia with properties that may be systematically different has obvious
important implications for cosmological use of SNe Ia, as it predicts
a redshift-dependent mix of the two types.  Since the use of SNe Ia
for cosmology requires correction of peak magnitudes from light-curve
shapes with a relation calibrated locally, the application of that
same relation to more distant SNe Ia may raise concerns.

\vspace{-0.18truein}
\acknowledgments
We gratefully acknowledge NASA support through {\it Chandra} Guest
Observer awards GO6-7061A, B, and C.  CB was supported by NASA through
Chandra Postdoctoral Fellowship Award PF6-70046 issued by the CXC,
operated by SAO for and on behalf of NASA under contract NAS8-03060.


\end{document}